\documentclass[12pt]{article}

\newcommand{\be}{\begin{equation}}
\newcommand{\ee}{\end{equation}}
\newcommand{\bi}[1]{\vspace{-3mm} \bibitem{#1}}

\newcommand{\et}{\rm \\ }

\newcommand{\bp}{{\bf Proof.} }
\newcommand{\ep}{$\ \ \ \Box$ \\ }

\usepackage{epsfig,amsmath,amssymb,graphics,graphicx} 

\usepackage{graphics,graphicx}
\usepackage{amssymb,amsmath}
\usepackage{hyperref}
\usepackage{amsfonts}
\usepackage{epsfig}

\begin{document}

\begin{center}

{\it ISRN Condensed Matter Physics. Vol.2014. (2014) 794097.} \\

\vskip 3mm
{\bf \large Fractional Gradient Elasticity from Spatial Dispersion Law } \\

\vskip 3mm
{\bf \large Vasily E. Tarasov} \\
\vskip 3mm

{\it Skobeltsyn Institute of Nuclear Physics,\\ 
Lomonosov Moscow State University, Moscow 119991, Russia} \\
{E-mail: tarasov@theory.sinp.msu.ru} \\

\begin{abstract}
Non-local elasticity models in continuum mechanics 
can be treated with two different approaches:
the gradient elasticity models (weak non-locality) and 
the integral non-local models (strong non-locality).
This article focuses on the fractional generalization 
of gradient elasticity that allows us to describe a weak non-locality of power-law type.
We suggest a lattice model with spatial dispersion of power-law type as
a microscopic model of fractional gradient elastic continuum. 
We demonstrate how the continuum limit transforms
the equations for lattice 
with this spatial dispersion into the continuum equations 
with fractional Laplacians in the Riesz's form.
A weak non-locality of power-law type in the non-local elasticity theory   
is derived from the fractional weak spatial dispersion in the lattice model.
The continuum equations with derivatives of non-integer orders, which are obtained from the lattice model, 
can be considered as a fractional generalization 
of the gradient elasticity.
These equations of fractional elasticity are solved for
some special cases: sub-gradient elasticity and super-gradient elasticity.
\end{abstract}

\end{center}

\noindent
PACS: 45.10.Hj; 61.50.Ah; 62.20.Dc \\


\section{Introduction}

The theory of derivatives and integrals of non-integer orders \cite{SKM,KST} 
allows us to investigate the behavior of materials and media 
that are characterized by non-locality of power-law type. 
Fractional calculus has a wide application in mechanics 
and physics (for example see \cite{CM} - \cite{IJMPB2013}).
Non-local elasticity theories in continuum mechanics 
can be treated with two different approaches \cite{AA2011}:
the gradient elasticity theory (weak non-locality) and 
the integral non-local theory (strong non-locality).
The fractional calculus allows us to formulate a
fractional generalization of non-local elasticity  models in two forms:
the fractional gradient elasticity models (weak power-law non-locality) and 
the fractional integral non-local models (strong power-law non-locality).
The idea to include some fractional integral term in 
the equations of the elasticity has been proposed by Lazopoulos in \cite{Laz}.
Fractional models of integral non-local elasticity are considered in different papers, 
see for example \cite{Laz,CCS-1,CCS-2,CCS-3,CPZ-1,CPZ-2,CPZ-3}.
The microscopic models of fractional integral elasticity are also described.
For this reason, the fractional integral elasticity models are not discussed here.

This article focuses on the fractional generalization 
of gradient elasticity which describes a weak non-locality of power type.
We suggest a lattice model with spatial dispersion of power-law type as
a microscopic model of fractional gradient elastic continuum. 
Complex lattice dynamics has been the subject of continuing interest in the theory of elasticity. 
As it was shown in \cite{JPA2006,JMP2006} (see also \cite{Chaos2006,LZ,CNSNS2006}), 
the equations with fractional derivatives can be directly connected to
lattice models with long-range interactions. 
In this paper, we consider models of lattices with spatial dispersion  
and its continuous limits. 
We define a map of lattice models into continuum models. 
A connection between the dynamics of lattice system 
of particles with long-range interactions 
and the fractional continuum equations is proved 
by using the transform operation \cite{JPA2006,JMP2006}. 
We make the transformation to the continuous limit 
and derive the fractional equation, which describes 
the dynamics of the non-local elastic materials.
We show how the continuous limit for the lattice with 
fractional weak spatial dispersion gives the corresponding 
continuum equation of the fractional gradient elasticity.
The continuum equations of fractional elasticity are solved for
some special cases: sub-gradient elasticity and super-gradient elasticity.

\section{Lattice Equations}

The lattice is characterized by space periodicity.
In an unbounded lattice we can define three non-coplanar vectors 
${\bf a}_1$, ${\bf a}_1$, ${\bf a}_1$, such that displacement 
of the lattice by the length of any of these vectors brings it back to itself. 
The vectors ${\bf a}_i$, $i=1,2,3$, are the shortest vectors 
by which a lattice can be displaced and be brought back into itself. 
As a result, all spatial lattice points 
can be defined by the vector ${\bf n} = (n_1,n_2,n_3)$, 
where $n_i$ are integer. 
If we choose the coordinate origin at one of the sites, then
the position vector of an arbitrary lattice site 
with ${\bf n} = (n_1,n_2,n_3)$ is written 
\be \label{Ko0}
{\bf r}({\bf n}) = \sum^3_{i=1} n_i {\bf a}_i  . \ee
In a lattice the sites are numbered in the same way as the particles, so 
that the vector ${\bf n}$ is at the same time "number vector" 
of a corresponding particle.

We assume that the equilibrium positions of particles coincide with 
the lattice sites ${\bf r}({\bf n})$.
A lattice site coordinate ${\bf r}({\bf n})$ differs from 
the coordinate of the corresponding particle, when particles 
are displaced relative to their equilibrium positions. 
To define the coordinates of a particle, it is necessary 
to indicate its displacement with respect to its equilibrium positions. 
We denote the displacement of a particle with vector ${\bf n}$ 
from its equilibrium position by the vector field ${\bf u} ({\bf n},t)$. 
 
The equation of motion of lattice particle is
\be \label{Ko2}
M \frac{\partial^2 u^k ({\bf n},t)}{\partial t^2}= 
- \sum_{ {\bf m} } K_{kl}({\bf n}, {\bf m} ) 
\, u^l ({\bf m} ,t) + F_k ( {\bf n}, t) , \ee
where $M$ is the mass of particle, 
$ F_k ( {\bf n}, t) $ are components of the external on-site force.
The italics $k,l$ are the coordinate indices. 
We assume the summation over doubly repeated 
coordinate indices from 1 to 3. 
The coefficients $K_{kl}({\bf n}, {\bf m} )$ describes 
the interparticle interaction in the lattice. 
For simplicity, we assume that all particles have the same mass $M$. 

It is easy see one important property of the coefficients 
$K_{kl}({\bf n}, {\bf m} )$. 
Assume the lattice to be displaced as a whole: 
$u^k({\bf n},t) = u^k =constant$. 
Then the internal lattice state cannot be changed 
in case of absence of external forces.
As a result, equation (\ref{Ko2}) gives 
\be \label{Ko21} 
\sum_{{\bf m} } K_{kl}({\bf n}, {\bf m} ) = 
\sum_{{\bf m} } K_{kl}({\bf m} , {\bf n}) = 0.  \ee
These conditions should be satisfied for any particle in the lattice, i.e., 
for any vector ${\bf n}$. 
Equations (\ref{Ko21}) follow from the conservation of total momentum in the lattice. 

For an unbounded homogeneous lattice, due to its homogeneity the matrix 
$K_{kl}({\bf n}, {\bf m} )$ has the form 
\be 
K_{kl}({\bf n}, {\bf m} ) = K_{kl}({\bf n} - {\bf m} ) , 
\ee
where elements of $K_{kl}({\bf n} - {\bf m} )$ of equation (\ref{Ko2})
are satisfied by the conditions 
\be \label{Ko22}
\sum_{ {\bf m} } K_{kl}({\bf n}- {\bf m} ) =
\sum_{{\bf n}} K_{kl}({\bf n}- {\bf m} ) = 0 . \ee
In a simple lattice each particle is an inversion center, and we have 
\be 
K_{kl}({\bf n}- {\bf m} ) = K_{kl}({\bf m} -{\bf n}) . 
\ee
Using condition (\ref{Ko22}), we can represent equations (\ref{Ko2}) in the form
\be \label{Kose1}
M \frac{\partial^2 u^k ({\bf n},t)}{\partial t^2}= 
- \sum_{ {\bf m} } K_{kl}({\bf n}, {\bf m} )  
\, \Bigl( u^l ({\bf n},t) - u^l ({\bf m} ,t) \Bigr) + 
F_k ({\bf n},t) . 
\ee
These equations of motion has the invariance 
with respect to its displacement of lattice as a whole
in case of absence of external forces 
even if the condition (\ref{Ko22}) is not satisfied. 
It should be noted that the noninvariant terms lead to the divergences 
in the continuous limit \cite{TarasovSpringer}.

Equation of motion (\ref{Kose1}) is equations for 
three-dimensional displacement vectors.
In this paper, we shall use the simplest model 
to describe the lattice, 
where all particles are displaced in one direction,
we assume that the displacement of particle from its equilibrium position 
is determined by a scalar rather than a vector. 
This model allows us to describe the main properties of 
the lattice using simple equations. 


The equations of motion for one-dimensional lattice system of 
interacting particles have the form
\be \label{2-Main_Eq_2}
M \, \frac{\partial^2 u_n(t)}{\partial t^2} = g 
\sum_{\substack{m=-\infty \\ m \ne n}}^{+\infty} \; K_{\alpha}(n,m) \; \Bigl( u_n(t)-u_m (t) \Bigr) 
 + F(n) ,
\ee
where we use the summation condition over repeated indexes.
Here $u_n(t)=u(n,t)$ are displacements from the equilibrium, 
$g$ is the coupling constant for interparticle interactions in the lattice, 
the term $F(n)$ characterizes an interaction of the particles   
with the external on-site force.

\section{Transform Operations for Lattice Equations}

Let us define the operation that transforms
the lattice equations for $u_n(t)$
into the continuum equation for a scalar field $u(x,t)$. 
In order to obtain continuum equation 
from the lattice equations, we assume 
that $u_n(t)$ are Fourier coefficients
of some function $\hat{u}(k,t)$.
We define the field $\hat{u}(k,t)$ on $[-k_0/2, k_0/2]$ by the equation 
\be \label{2-ukt}
\hat{u}(k,t) = \sum_{n=-\infty}^{+\infty} \; u_n(t) \; e^{-i k x_n} =
{\cal F}_{\Delta} \{u_n(t)\} ,
\ee
\be \label{2-un} 
u_n(t) = \frac{1}{k_0} \int_{-k_0/2}^{+k_0/2} dk \ \hat{u}(k,t) \; e^{i k x_n}= 
{\cal F}^{-1}_{\Delta} \{ \hat{u}(k,t) \} ,
\ee
where $x_n = n d$, and $d=2\pi/k_0$ is the inter-particle distance. 
For simplicity, we assume that all particles have the same inter-particle distance $d$.
Equations (\ref{2-ukt}) and (\ref{2-un}) can be used to obtain 
the Fourier transform in the limit $d \to 0$ ($k_0 \to \infty$). 
Then change the sum to an integral, and 
equations (\ref{2-ukt}) and (\ref{2-un}) become
\be \label{2-ukt2} 
\tilde{u}(k,t)=\int^{+\infty}_{-\infty} dx \ e^{-ikx} u(x,t) = 
{\cal F} \{ u(x,t) \}, 
\ee
\be \label{2-uxt}
u(x,t)=\frac{1}{2\pi} \int^{+\infty}_{-\infty} dk \ e^{ikx} \tilde{u}(k,t) =
 {\cal F}^{-1} \{ \tilde{u}(k,t) \} . 
\ee
We replace the discrete function
\be 
u_n(t) = \frac{2 \pi}{k_0} u(x_n,t) 
\ee 
by continuous field $u(x,t)$ considering $x_n=nd= 2 \pi n /k_0 \to x $. 
We assume that $\tilde{u}(k,t)= {\cal L} \hat{u}(k,t)$,
where ${\cal L}$ denotes the passage 
to the limit $d \to 0$ ($k_0 \to \infty$).
Here $\tilde{u}(k,t)$ is a Fourier transform of the field $u(x,t)$,
and $\hat{u}(k,t)$ is a Fourier series transform of $u_n(t)$,
where we use $u_n(t)=(2\pi/k_0) u(nd,t)$.
The function $\tilde{u}(k,t)$ can be derived from $\hat{u}(k,t)$
in the limit $d \to 0$.

As a result, we define
the map from a lattice model into a continuum model 
by the transform operation $\hat T$, which is the combination \cite{JPA2006,JMP2006}  
$\hat T={\cal F}^{-1} {\cal L} \ {\cal F}_{\Delta}$ 
of the following operations: 
\begin{enumerate} 
\item
The Fourier series transform:
\be \label{2-O1}
{\cal F}_{\Delta}: \quad u_n(t) \to {\cal F}_{\Delta}\{ u_n(t)\}=
\hat{u}(k,t) .
\ee
\item 
The passage to the limit $d \to 0$:
\be 
{\cal L}: \quad \hat{u}(k,t) \to {\cal L} \{\hat{u}(k,t)\}=
\tilde{u}(k,t) . \ee
\item 
The inverse Fourier transform: 
\be
{\cal F}^{-1}: \quad \tilde{u}(k,t) \to 
{\cal F}^{-1} \{ \tilde{u}(k,t)\}=u(x,t) .
\ee
\end{enumerate}

The similar transformations can be performed 
for differential equations to map the lattice equation into an equation for the elastic continuum.
Therefore the operation 
$\hat T={\cal F}^{-1} {\cal L} \ {\cal F}_{\Delta}$ 
allows us to realize transformation of lattice models 
of interacting particles into continuum models 
\cite{JPA2006,JMP2006}.


Let us consider the Fourier series transform of the interaction term. 

{\bf Proposition 1.} {\it 
Let $K_{\alpha}(n,m)$ be such that the conditions
\be \label{2-Knm1}
K_{\alpha}(n,m)=K_{\alpha}(n-m)=K_{\alpha}(m-n) ,  \quad
\sum^{\infty}_{n=1} |K_{\alpha}(n)|^2 < \infty  
\ee
hold. Then the Fourier series transform ${\cal F}_{\Delta}$ 
maps the term
\be \label{2-C1-L}
\sum^{+\infty }_{\substack{m=-\infty \\ m \not= n}}
K_{\alpha}(n,m) \, \Bigl( u_n(t)-u_m(t) \Bigr) ,
\ee
where $u_n=u_n(t)$ is a position of the $n$th particle,
into the term 
\be \label{2-20}
{\cal F}_{\Delta} \left( \sum^{+\infty }_{\substack{m=-\infty \\ m \not= n}}
K_{\alpha}(n,m) \, \Bigl( u_n(t)-u_m(t) \Bigr) \right) =
\Bigl( \hat{K}_{\alpha}(0)- \hat{K}_{\alpha}(k d) \Bigr) \, \hat u(k,t) ,
\ee 
where }
\be 
\hat{K}_{\alpha}(k d)={\cal F}_{\Delta}\{ K_{\alpha}(n)\} , 
\quad  \hat{u}(k,t)={\cal F}_{\Delta}\{ u_n(t)\} . 
\ee

\bp
To derive the Fourier series transform of the interaction term (\ref{2-C1-L}), 
we multiply (\ref{2-C1-L}) by $\exp(-ikn d)$, 
and sum over $n$ from $-\infty$ to $+\infty$. Then
\[ \sum^{+\infty}_{n=-\infty} \
\sum^{+\infty}_{\substack{m=-\infty \\ m \not=n}}
e^{-ikn d}  K_{\alpha}(n-m) \ \Bigl( u_n-u_m \Bigr) = \]
\be \label{2-C6}
=\sum^{+\infty}_{n=-\infty} \  \sum^{+\infty}_{\substack{m=-\infty \\ m \not=n}}
e^{-ikn d} K_{\alpha}(n-m) u_n - 
\sum^{+\infty}_{n=-\infty} \sum^{+\infty}_{\substack{m=-\infty \\ m \not=n}} 
e^{-ikn d} K_{\alpha}(n-m) u_m .
\ee
Using the conditions (\ref{2-Knm1}), 
we introduce the notations
\be \label{2-not}
\hat{K}_{\alpha}(k d)=
\sum^{+\infty}_{\substack{n=-\infty \\ n\not=0}} 
e^{-ikn d} K_{\alpha}(n) ,
\ee
\be \label{2-C4}
\hat u(k,t)=\sum^{+\infty}_{n=-\infty} e^{-ikn d} u_n(t) .
\ee
Using $K_{\alpha}(-n)=K_{\alpha}(n)$, the function (\ref{2-not}) can be represented by
\be \label{2-Kcos}
\hat{K}_{\alpha}(k d) =\sum^{+\infty}_{n=1} K_{\alpha}(n) 
\left( e^{-iknd} +e^{iknd} \right) = 
2\sum^{+\infty}_{n=1} K_{\alpha}(n) \cos \left( k d \right) .
\ee
From equation (\ref{2-Kcos}), we can see that $\hat{K}_{\alpha}(k d)$ 
is a periodic function
\be 
\hat{K}_{\alpha}(k d+2\pi m)= \hat{K}_{\alpha}(k d ) , 
\ee
where $m$ is an integer. 
Using (\ref{2-C4}) and (\ref{2-not}), 
the first term on the right-hand side of (\ref{2-C6}) gives
\be \label{2-C7} 
\sum^{+\infty}_{n=-\infty} \ \sum^{+\infty}_{\substack{m=-\infty \\ m \not=n}}
e^{-ikn d} K_{\alpha}(n-m) u_n =
\sum^{+\infty}_{n=-\infty} e^{-ikn d} u_n 
\sum^{+\infty}_{\substack{m^{\prime}=-\infty \\ m^{\prime} \not=0}}
K_{\alpha}(m^{\prime})= \hat u(k,t) \hat{K}_{\alpha}(0) .
\ee
Here we use (\ref{2-Knm1}), and $K_{\alpha}(m^{\prime}+n-n)=K_{\alpha}(m^{\prime})$, and
\be \label{2-C8}
\hat{K}_{\alpha}(0)=\sum^{+\infty}_{\substack{n=-\infty \\ n \not=0}}
K_{\alpha}(n)=2\sum^{\infty}_{n=1} K_{\alpha}(n) .
\ee
Using $K_{\alpha}(m,n^{\prime}+m)=K_{\alpha}(n^{\prime})$,
the second term on the right-hand side of (\ref{2-C6}) has the form
\[ \sum^{+\infty}_{n=-\infty} \ 
\sum^{+\infty}_{\substack{m=-\infty \\ m \not=n}}
e^{-ikn d} K_{\alpha}(n-m) u_m = 
\sum^{+\infty}_{m=-\infty} u_m 
\sum^{+\infty}_{\substack{n=-\infty \\ n \not=m}} 
e^{-ikn d} K_{\alpha}(n-m) = \]
\be \label{2-C9}
=\sum^{+\infty}_{m=-\infty } u_m e^{-ikm d}
\sum^{+\infty}_{\substack{n^{\prime}=-\infty \\ n^{\prime}\not=0}} 
e^{-ikn^{\prime} d} K_{\alpha}(n^{\prime})=
\hat u(k,t)\hat{K}_{\alpha}(k d) .
\ee
Equation (\ref{2-C7}) and (\ref{2-C9}) give the expression
\be 
\Bigl( \hat{K}_{\alpha}(0)- \hat{K}_{\alpha}(k d) \Bigr) \, \hat u(k,t) ,
\ee 
where $\hat{K}_{\alpha}(k d)$ is 
defined by equation (\ref{2-not}). 
\ep

Let us give the statement that describes 
the Fourier transform of the lattice equations.

{\bf Proposition 2.} {\it 
The Fourier series transform ${\cal F}_{\Delta}$ 
maps the lattice equations of motion 
\be \label{2-C1}
M \, \frac{\partial^2 u_n(t)}{\partial t^2}=g              
\sum^{+\infty }_{\substack{m=-\infty \\ m \not= n}}
K_{\alpha}(n-m) \, \Bigl( u_n(t)-u_m(t) \Bigr) + F(n) ,
\ee
where $K_{\alpha}(n-m)$ satisfies the conditions (\ref{2-Knm1}), 
into the continuum equation 
\be \label{2-20eq}
M \, \frac{\partial^2  \hat u(k,t)}{\partial t^2}=
g \Bigl( \hat{K}_{\alpha}(0)- \hat{K}_{\alpha}(k d) \Bigr) \, \hat u(k,t) 
+ {\cal F}_{\Delta} \{F(n)\} ,
\ee 
where   $\hat{u}(k,t)={\cal F}_{\Delta}\{ u_n(t)\}$,  \ 
$\hat{K}_{\alpha}(k d)={\cal F}_{\Delta}\{ K_{\alpha}(n)\}$, 
and ${\cal F}_{\Delta}$ is an operator notation for the Fourier
series transform.}

\bp
To derive the equation for the field $\hat u(k,t)$, we
multiply equation (\ref{2-C1}) by $\exp(-ikn d)$, 
and sum over $n$ from $-\infty$ to $+\infty$. Then
\be \label{2-C3a-E}
\sum^{+\infty}_{n=-\infty} e^{-ikn d} 
\frac{\partial^2}{\partial t^2}u_n(t)= 
g  \sum^{+\infty}_{n=-\infty} \
\sum^{+\infty}_{\substack{m=-\infty \\ m \not=n}}
e^{-ikn d}  K_{\alpha}(n-m) \ \Bigl( u_n-u_m \Bigr) +
\sum^{+\infty}_{n=-\infty} e^{-iknd} F(n) .
\ee
Using  (\ref{2-C4}) the left-hand side of (\ref{2-C3a-E}) has the form
\be 
\sum^{+\infty}_{n=-\infty} e^{-ikn d} 
\frac{\partial^2 u_n(t)}{\partial t^2}=
\frac{\partial^2 }{\partial t^2}
\sum^{+\infty}_{n=-\infty} e^{-ikn d} u_n(t)=
\frac{\partial^2 \hat u(k,t)}{\partial t^2} . 
\ee
The second term of the right-hand side of (\ref{2-C3a-E}) is
\be 
\sum^{+\infty}_{n=-\infty} e^{-ikn d} F(n)=
{\cal F}_{\Delta} \{F(n)\} . 
\ee
The Fourier series transform ${\cal F}_{\Delta}$ 
maps the interaction term (\ref{2-C1-L}) into expression (\ref{2-20}).
As a result, we obtain equation (\ref{2-C3a-E}) 
in the form (\ref{2-20eq}),
where ${\cal F}_{\Delta} \{F(n)\}$ is an operator notation 
for the Fourier series transform of $F(n)$. 
\ep

\section{Fractional Weak Spatial Dispersion}

\subsection{Weak Spatial Dispersion}

Spatial dispersion is the dependence of $\hat{K}_{\alpha}(|{\bf k}|)$
on the wave vector ${\bf k}$ that leads to non-local properties of the continuum.
The spatial dispersion gives non-local connection 
between the stress tensor $\sigma_{kl}$ and the strain tensor $\varepsilon_{kl}$. 
The tensor $\sigma_{kl}$ at any point ${\bf r}$ of the continuum is not uniquely defined by
the values of $\varepsilon_{kl}$ at this point. 
It also depends on the values of $\varepsilon_{kl}$ 
at neighboring points ${\bf r}^{\prime}$, located near the point ${\bf r}$.

A non-local constitutive relation 
between the stress $\sigma_{kl}$ and the strain 
$\varepsilon_{kl}$ can be understood 
on the basis of analysis of a lattice model.
The particles of the lattice 
oscillate about their equilibrium positions and interact with each other.
The equations of oscillations of the lattice particles
with the local (nearest-neighbor) interaction   
gives the partial differential equation of integer orders
in the continuum limit \cite{JPA2006,JMP2006}. 
Note that the lattice with non-local (long-range) interactions  
in the continuous limit can give 
fractional partial differential equations 
for non-local continuum \cite{JPA2006,JMP2006}.

Qualitatively describing the process we can say that
the fields of the elastic wave moves particles from their equilibrium positions 
at a given point ${\bf r}$, which causes an additional shift of the particles
in neighboring and more distant points ${\bf r}^{\prime}$ in some neighborhood.
Therefore, the properties of the medium, and hence the stress tensor field $\sigma_{kl}$ depends 
on the values of strain tensor field $\varepsilon_{kl}$ not only in a selected point, 
but also in its neighborhood.

The size of the area in which the kernel $\hat{K}_{\alpha}(|{\bf k}|)$
is significant is determined by the characteristic length of
interaction $R_0$. 
The size $R_0$ of the area is usually of the order of the lattice constant. 
Wavelength $\lambda$ of elastic wave is several orders larger than the size of this region, 
so the values of the field of elasticity wave do not change 
for a region of size $R_0$.
By other words, the wavelength $\lambda$ usually holds 
$k R_0 \sim R_0 / \lambda \ll 1$. 
In such lattice the spatial dispersion is weak. 
To describe the lattice dynamics it is enough to know the dependence of 
the function $\hat{K}_{\alpha}(|{\bf k}|)$ only for small values 
$k=|{\bf k}|$ and we can replace this function by 
the Taylor's polynomial series.
For an isotropic linear medium, we use 
\be \label{Tay1}
\hat{K}_{\alpha}(k) = \hat{K}_{\alpha}(0) + a_1 \, k + a_2 \, k^2 + ... \quad .
\ee
Here we neglect a frequency dispersion, and so $\hat{K}_{\alpha}(0)$, $a_1$, $a_2$ 
do not depend on the frequency $\omega$.


\subsection{Fractional Taylor series approach}

The weak spatial dispersion in the media with
power-law type of non-locality cannot be describes by
the usual Taylor approximation. 
The fractional Taylor series is very useful for approximating 
non-integer power-law functions \cite{AP2013}. 
For example, the usual Taylor series for the non-linear power-law function
\be \label{ve}
\hat{K}_{\alpha}(k) = a_0 + a_{\alpha} k^{\alpha}  
\ee
has infinitely many terms for non-integer $\alpha$.

If we use the fractional Taylor's formula (see Appendix 1)
we get finite number of terms.
For example, the Taylor's series in the Odibat-Shawagfeh form
that contains the Caputo fractional derivative $_a^CD^{\alpha}_k$ has two terms for (\ref{ve}).
Using 
\be \label{T2}
_0^CD^{\alpha}_k k^{\beta} = \frac{\Gamma(\beta+1)}{\Gamma(\beta-\alpha+1)} \, k^{\beta-\alpha} , 
\quad (k>0, \ \alpha>0, \ \beta >0)
\ee
for the case $\beta=\alpha$, we get
\be
\, _0^CD^{\alpha}_k k^{\alpha} = \Gamma(\alpha+1), \quad (\, _0^CD^{\alpha}_k )^n k^{\alpha} =0 . 
\ee
As a result, we have
\be 
(\, _0^CD^{\alpha}_k \hat{K}_{\alpha})(0) = \Gamma(\alpha+1), \quad
( (\, _0^CD^{\alpha}_k)^n \hat{K}_{\alpha})(0) = 0, \quad (n \ge 2)
\ee
and the fractional Taylor's series approximation of function (\ref{ve}) is exact.

\subsection{Weak spatial dispersion of power-law types}

We consider properties of the lattice with weak spatial dispersion 
that is described by the function $\hat{K}_{\alpha}(|{\bf k}|)$ 
of a non-integer power-law type.
In the continuous limit this model gives a model of 
continuum with power-law non-locality. 

The Fourier series transform ${\cal F}_{\Delta}$ of 
the interaction term (\ref{2-C1-L}) is defined by (\ref{2-20}), where 
\be \label{2-Kak}
\hat{K}_{\alpha}(|{\bf k}|)=\sum^{+\infty}_{\substack{n=-\infty \\ n\not=0}} 
e^{-ikn} K_{\alpha}(n) = 2 \sum^{\infty}_{n=1} K_{\alpha}(n) \cos(n |{\bf k}|) ,
\ee
and $\hat u(k,t)={\cal F}_{\Delta}\{u_n(t)\}$.
If the function $\hat{K}_{\alpha}(|{\bf k}|)$ is given, then
$K_{\alpha}(n)$ can be defined by
\be \label{2-Knn}
K_{\alpha}(n)=\frac{1}{\pi} \int^{\pi}_{0} \hat{K}_{\alpha}(|{\bf k}|) \cos(n |{\bf k}|) \ d |{\bf k}| .
\ee

The weak spatial dispersion will be called $\alpha_1$-type,  
if the function (\ref{2-Kak}) satisfies the condition
\be \label{ea1}
\lim_{|{\bf k}| \to 0} 
\frac{\hat{K}_{\alpha}(|{\bf k}|) - \hat{K}_{\alpha}(0) }{ |{\bf k}|^{\alpha_1}} =a_{\alpha_1},
\ee
where $\alpha_1>0$ and $0<|a_{\alpha_1}|< \infty$.
The weak spatial dispersion (and the interparticle interaction in the lattice)
will be called $\alpha=(\alpha_1,\alpha_2)$-type,  
if the function  $\hat{K}_{\alpha}(|{\bf k}|)$ satisfies the conditions (\ref{ea1}) and
\be \label{ea2}
\lim_{|{\bf k}| \to 0} 
\frac{ \hat{K}_{\alpha}(|{\bf k}|) - \hat{K}_{\alpha}(0) - 
a_{\alpha_1} \, |{\bf k}|^{\alpha_1} }{ |{\bf k}|^{\alpha_2}} = a_{\alpha_2},
\ee
where $\alpha_2>\alpha_1>0$ and $0<|a_{\alpha_2}|< \infty$.

Similarly we define the weak spatial dispersion and the interaction in the lattice
of the $\alpha=(\alpha_1, . . . , \alpha_N)$-type.
For the weak spatial dispersion of the $\alpha=(\alpha_1, . . . , \alpha_N)$-type,   
the function  $\hat{K}_{\alpha}(|{\bf k}|)$ can be represented in the form
\be
\hat{K}_{\alpha}(|{\bf k}|)= \hat{K}_{\alpha}(0) + 
\sum^N_{j=1} a_{\alpha_j} |{\bf k}|^{\alpha_j} + R^{(N)}_{\alpha}(|{\bf k}|) ,
\ee
where $0<\alpha_1< \alpha_2 < . . . < \alpha_N$, and
\be 
\lim_{|{\bf k}| \to 0} 
\frac{ R^{(N)}_{\alpha}(|{\bf k}|)}{|{\bf k}|^{\alpha_N}} =0.
\ee
As a result, we can use the following approximation for
weak spatial dispersion 
\be \label{approx-1}
\hat{K}_{\alpha}(|{\bf k}|) \approx \hat{K}_{\alpha}(0) + 
\sum^N_{j=1} a_{\alpha_j} |{\bf k}|^{\alpha_j} .
\ee

If $\alpha_j=j$ for all $j \in \mathbb{N}$, we can use the usual Taylor's formula. 
In this case we have the usual case of the weak spatial dispersion.
In general, we should use a fractional generalization of the Taylor's series (see Appendix 1). 
If the orders of the fractional Taylor series approximation 
will be correlated with the type of weak spatial dispersion, 
then the fractional Taylor series approximation of  $\hat{K}_{\alpha}(|{\bf k}|)$ will be exact. 
In the general case $0<\alpha_{j+1}- \alpha_j<1$, we can use the fractional Taylor's formula 
in the Dzherbashyan-Nersesian form (see Appendix 1). For the special cases $\alpha_j= j \, \alpha_1$, 
where $\alpha_1<1$ and/or $\alpha_j=\alpha+j$, 
we could use other kind of the fractional Taylor's formulas.


\section{Fractional Gradient Elasticity Equation for Continuum}

In the continuous limit the equation for lattice 
with the interaction of the $\alpha$-type
gives the equation for continuum of the fractional gradient model.

{\bf Proposition 3.} 
{\it In the continuous limit the lattice equation of motion
\be \label{2-54b}
M \, \frac{\partial^2 u_n(t)}{\partial t^2} = 
g  \sum_{\substack{m=-\infty \\ m \ne n}}^{+\infty} \; 
K_{\alpha}(n-m) \; \Bigl( u_n(t) -u_m(t) \Bigr) + F (n) 
\ee
with the weak spatial dispersion of the $\alpha$-type
gives the fractional continuum equation of the form 
\be \label{2-CME}
\frac{\partial^2 u(x,t)}{\partial t^2} =
- \sum^N_{j=1} G_{\alpha_j} \, ((-\Delta)^{\alpha_j/2} u) (x,t) +
\frac{1}{\rho} \, f(x) ,
\ee
where $(-\Delta)^{\alpha_j/2}$ is the fractional Laplacian of order $\alpha_j$ 
in the Riesz's form (see Appendix 2), the variables $x$ and $d=d$ are dimensionless,
$f(x)=F(x)/(A d)$, $\rho= M / (A d)$, and 
\be \label{Gj} 
G_{\alpha_j}= \frac{g \, a_{\alpha_j} \, d^{\alpha_j}}{M}  \quad (j=1, . . . , N) 
\ee
are finite parameters.
\et

\bp
The Fourier series transform ${\cal F}_{\Delta}$ of equation (\ref{2-54b})
gives (\ref{2-20eq}).
After division by the cross-section area of the medium $A$ 
and the inter-particle distance $d$, 
the limit $d \to 0$ for equation (\ref{2-20eq}) gives
\be \label{2-Eq-k}
\frac{\partial^2}{\partial t^2} \hat{u}(k,t) =
\sum^N_{j=1} \frac{g \, d^{\alpha_j}}{M} \; \hat{\mathcal{K}}_{\alpha_j, \Delta}(k) \; \hat{u}(k,t)  
+ \frac{1}{\rho} \mathcal{F}_{\Delta} \{ f (n) \}  , 
\ee
where $\rho= M/ (A d)$ is the mass density, 
$d$ is the inter-particle distance, $f(n) = F(n)/ (A d)$, and
\be
\hat{\mathcal{K}}_{\alpha_j, \Delta}(k) = - a_{\alpha_j} |k|^{\alpha_j}  
-R^{(N)}_{\alpha} (k d) d^{-\alpha_j} . \ee
Here we use (\ref{approx-1}), and $G_{\alpha_j}$ ($j=1, . . . ,N$) 
are finite parameters that are defined by (\ref{Gj}). 
Note that $R^{(N)}_{\alpha}$ satisfies the condition
\be
\lim_{d \to 0} 
\frac{R^{(N)}_{\alpha} (k d)}{d^{\alpha_N}} =0 .
\ee
The expression for $\hat{\mathcal{T}}_{\alpha_j,\Delta} (k)$ can be considered
as a Fourier transform of the interaction term (see Proposition 1). 
Note that $g a_{\alpha_j}  \to \infty$
for the limit $d \to 0$, if $G_{\alpha_j}$ are finite parameters.

In the limit $d \to 0$, equation (\ref{2-Eq-k}) gives
\be \label{2-Eq-k2}
\frac{\partial^2 \tilde{u}(k,t)}{\partial t^2} =
\sum^N_{j=1} G_{\alpha_j} \; \hat{\mathcal{T}}_{\alpha_j}(k) \; \tilde{u}(k,t)  
+ \frac{1}{\rho} \mathcal{F} \{ f (x) \}  , 
\ee
where
\be 
\hat{\mathcal{K}}_{\alpha_j}(k) =
{\cal L}\hat{\mathcal{K}}_{\alpha_j, \Delta}(k) 
= - a_{\alpha_j} |k|^{\alpha_j} ,  \quad
\tilde{u}(k,t)={\cal L} \hat{u}(k,t) . 
\ee
The inverse Fourier transform of (\ref{2-Eq-k2}) has the form
\be \label{2-Eq-x}
\frac{\partial^2 u(x,t)}{\partial t^2} =
\sum^N_{j=1} G_{\alpha_j} \; \mathcal{T}_{\alpha_j}(x) \; u(x,t) + 
\frac{1}{\rho} \, f (x) ,
\ee
where 
\be \label{2-Tx0}
\mathcal{T}_{\alpha_j}(x) =  \mathcal{F}^{-1} \{ \hat{\mathcal{K}}_{\alpha_j} (k) \} = 
- a_{\alpha_j} (-\Delta)^{\alpha_j/2}   .
\ee
Here, we use the connection between the Riesz fractional 
Laplacian $(-\Delta)^{\alpha/2}$ 
and its Fourier transform (see Appendix 2 and \cite{SKM,KST}): 
\be \label{FFL}
{\cal F}[ (-\Delta)^{\alpha/2} u({\bf r})]({\bf k})= 
|{\bf k}|^{\alpha} \, \hat u({\bf k})
\ee
in the form
\be 
|k|^{\alpha_j} \longleftrightarrow (-\Delta)^{\alpha_j/2} . 
\ee
Substitution of (\ref{2-Tx0}) into (\ref{2-Eq-x}) gives
the continuum equation (\ref{2-CME}). 
\ep

Equations (\ref{2-CME}) and (\ref{Gj}) give the close relation between 
the discrete microstructure of lattice with weak spatial dispersion
of power-law type and the fractional gradient models of weak non-local continuum. 

Let us consider the special case $\alpha_j=j$ for integer $j \in \mathbb{N}$. 
If the function $\hat{K}_{\alpha}(k)$ has the form
\be \label{a2}
\hat{K}_{\alpha}(k) \approx \hat{K}_{\alpha}(0) + a_2 \, k^2 ,
\ee
then we get the well-known equation
\be 
\frac{\partial^2 u(x,t)}{\partial t^2} = G_2 \, \Delta u(x,t) + \frac{1}{\rho} \, f(x) .
\ee
Here 
\be
G_2 = \frac{g \, a_2 \, d^2}{M \, A} = \frac{E}{\rho} , 
\ee
where $E=K d/A$ is the Young's modulus,
$K= g a_2$ is the spring stiffness,  $\rho= M/ (A d)$ is the mass density.

If we can use the spatial dispersion law in the form
\be \label{a2b}
\hat{K}_{\alpha}(k) \approx \hat{K}_{\alpha}(0) + a_2 \, k^2 + a_4 \, k^4 ,
\ee
then we have the equation of the gradient elasticity as
\be 
\frac{\partial^2 u(x,t)}{\partial t^2} = G_2 \, \Delta u(x,t) - 
G_4 \, \Delta^2 u(x,t)  +  \frac{1}{\rho} \, f(x) ,
\ee
where $\alpha_j=j$, 
\be 
G_4 = \frac{g \, a_4 \, d^4}{M \, A} = \frac{a_4 \, E \, d^2}{a_2 \, \rho} , \quad 
a_j = \left( \frac{\partial^j \hat{K}_{\alpha}(k)}{\partial k^j}\right)_{k=0} . 
\ee
The scale parameter $l^2$ of the gradient elasticity is connected with
the coupling constants of the lattice by the equation
\be \label{L2}
l^2 =\frac{ \left| a_4 \right| \, d^2}{|a_2|} .
\ee
The second-gradient term is preceded by the sign that is defined by
$\operatorname{sgn} (a_4/a_2)$. 

Similarly, we can consider more general model of lattice 
with fractional weak spatial dispersion of $\alpha=(\alpha_1, . . . , \alpha_N)$-type
\be  \label{peps-G}
\hat{K}_{\alpha}({\bf k}) = \hat{K}_{\alpha}(0) + \sum^N_{j=1} a_{\alpha_j} \, |{\bf k}|^{\alpha_j} . 
\ee
Then the continuum equation for fractional gradient model has the form
\be \label{2-CME-3}
\frac{\partial^2 u({\bf r},t)}{\partial t^2} =
- \sum^N_{j=1} c_j \, ((-\Delta)^{\alpha_j/2} u) ({\bf r},t) +
\frac{1}{\rho} \, f({\bf r}) ,
\ee
where we use new notation for the constants, $c_j=G_{\alpha_j}$.
Note that ${\bf r}$ and $r=|{\bf r}|$ are dimensionless.


\section{Solution of Fractional Gradient Elasticity Equation}

\subsection{Plane wave solution}

Let us consider the plane waves $u({\bf r},t) = e^{- i \omega \, t} \, u({\bf r})$. 
Then equation (\ref{2-CME-3}) gives
\be \label{FPDE-1}
\sum^N_{j=1} c_j \, ((-\Delta)^{\alpha_j/2} u) ({\bf r}) - \omega^2 u({\bf r}) =
\frac{1}{\rho} f({\bf r}) .
\ee

We apply the Fourier method to solve fractional equation (\ref{FPDE-1}), 
which is based on the relation 
\be \label{FFL2}
{\cal F}[ (-\Delta)^{\alpha/2} u({\bf r})]({\bf k})= 
|{\bf k}|^{\alpha} \, \hat u({\bf k}).
\ee
Applying the Fourier transform ${\cal F}$ to both sides of (\ref{FPDE-1}) and using (\ref{FFL2}), we have
\be
({\cal F} u)({\bf k}) = \frac{1}{\rho} \,
\left( \sum^N_{j=1} c_j |{\bf k}|^{\alpha_j} - \omega^2 \right)^{-1} 
({\cal F} f)({\bf k}) .
\ee

The fractional analog of the Green function (see Section 5.5.1. in \cite{KST}) is given by
\be \label{FGF}
G^n_{\alpha}({\bf r})= {\cal F}^{-1} \Bigl[ \left( \sum^N_{j=1} c_j |{\bf k}|^{\alpha_j} -\omega^2 \right)^{-1} \Bigr] ({\bf r})=
\int_{\mathbb{R}^n} \left( \sum^N_{j=1} c_j |{\bf k}|^{\alpha_j} - \omega^2 \right)^{-1} \
e^{ + i ({\bf k},{\bf r}) } \, d^n {\bf k} ,
\ee
where $\alpha=(\alpha_1,...,\alpha_m)$.

The following relation
\be \label{3-1}
\int_{\mathbb{R}^n} e^{  i ({\bf k},{\bf r}) } \, f(|{\bf k}|) \, d^n {\bf k}= 
\frac{(2 \pi)^{n/2}}{ |{\bf r}|^{(n-2)/2}} 
\int^{\infty}_0 f( \lambda) \, \lambda^{n/2} \, J_{n/2-1}(\lambda |{\bf r}|) \, d \lambda
\ee
holds (see Lemma 25.1 of \cite{SKM}) for any suitable function $f$
such that the integral in the right-hand side of (\ref{3-1}) is convergent. 
Here $J_{\nu}$ is the Bessel function of the first kind. 
As a result, the Fourier transform of a radial function is also a radial function.


Using relation (\ref{3-1}), the Green function (\ref{FGF}) 
can be represented (see Theorem 5.22 in \cite{KST})
in the form of the integral with respect to one parameter $\lambda$
\be \label{G-1}
G^n_{\alpha} ({\bf r}) = \frac{|{\bf r}|^{(2-n)/2}}{(2 \pi)^{n/2}} 
\int^{\infty}_0 \left( \sum^N_{j=1} c_j \lambda^{\alpha_j} - \omega^2 \right)^{-1} 
\lambda^{n/2} \, J_{(n-2)/2} (\lambda |{\bf r}|) \, d \lambda,
\ee
where $n=1,2,3$ and $\alpha=(\alpha_1,...,\alpha_m)$, and 
$J_{(n-2)/2}$ is the Bessel function of the first kind .

For the 3-dimensional case, we use 
\be
J_{1/2} (z) = \sqrt{\frac{2}{\pi z}} \, \sin (z).
\ee
Then we have
\be \label{G-1-3D}
G^3_{\alpha} ({\bf r}) =\frac{1}{ 2 \pi^2 |{\bf r}| } 
\int^{\infty}_0 \left( \sum^N_{j=1} c_j \lambda^{\alpha_j} - \omega^2 \right)^{-1} 
\, \lambda \, \sin (\lambda |{\bf r}|) \, d \lambda .
\ee
For the 1-dimensional case, we use 
\be
J_{-1/2} (z) = \sqrt{\frac{2}{\pi z}} \, \cos (z).
\ee
Then we have (see Theorem 5.24 in \cite{KST} pages 345-346) the function
\be 
G^1_{\alpha} ({\bf r}) =\frac{1}{\pi} 
\int^{\infty}_0 \left( \sum^m_{s=1} a_{\alpha_j} \lambda^{\alpha_j} - \omega^2 \right)^{-1} 
\, \cos (\lambda |{\bf r}|) \, d \lambda .
\ee

If $\alpha_N > 1$ and $c_N \ne 0$, then equation (\ref{FPDE-1}) 
(see, for example, Section 5.5.1. pages 341-344 in  \cite{KST}) 
has a particular solution $u(|{\bf r}|)$. 
Such particular solution is represented in the form of 
the convolution of the functions $G^n_{\alpha}(|{\bf r}|)$ and $f(|{\bf r}|)$ as follow
\be \label{phi-G}
u ({\bf r})=  \frac{1}{\rho} \, \int_{\mathbb{R}^n} G^n_{\alpha} ({\bf r} - {\bf r}^{\prime}) \, 
f ({\bf r}^{\prime}) \, d^n {\bf r}^{\prime} ,
\ee
where the Green function $G^n_{\alpha}({\bf r})$ is given by (\ref{G-1}). 

In 3-dimensional case the function $f(|{\bf r}|)$ does not depend on the angles.
Therefore we can use the spherical coordinates and then reduce the integration 
$d^3 {\bf r}^{\prime}$ in (\ref{phi-G}) to $dr=d|{\bf r}|$
by integrating with respect to the angles
\be \label{Nano}
u (r)=  \frac{4 \pi}{\rho} \, \int_{\mathbb{R}} G^3_{\alpha} (|{\bf r} - {\bf r}^{\prime}|) \, 
f (r^{\prime}) \, (r^{\prime})^2 \, d r^{\prime} ,
\ee
where $r=|{\bf r}|$ and $r^{\prime}=|{\bf r}^{\prime}|$.

\subsection{Static solution}

Let us consider the statics ($\partial u({\bf r},t) /\partial t =0$, i.e. $u({\bf r},t) = u({\bf r})$) 
in the suggested fractional gradient elasticity model.
We can consider the fractional partial differential equation (\ref{FPDE-1}) 
with $\omega^2=0$ and $c_1 \ne 0$, when $N \ge 1$, and also 
the case where $\alpha_1< 3$, $\alpha_N > 1$, $N \ge 1$, $c_1 \ne 0$, 
$c_N \ne 0$, $\alpha_N > ... > \alpha_1>0$, which is given by
\be \label{FPDE-3}
\sum^N_{j=1} c_j \, ((-\Delta)^{\alpha_j/2} u) ({\bf r}) = \frac{1}{\rho} \, f({\bf r}) .
\ee
Equation (\ref{FPDE-3}) has the following particular solution (see Theorem 5.23 in \cite{KST}), 
that is represented in the form of the convolution of the functions as
\be \label{phi-G3}
u ({\bf r})= \frac{1}{\rho} \, \int_{\mathbb{R}^n} G^n_{\alpha} ({\bf r} - {\bf r}^{\prime}) \, 
f ({\bf r}^{\prime}) \, d^n {\bf r}^{\prime} 
\ee
with the Green function 
\be \label{G-3}
G^n_{\alpha} ({\bf r}) = \frac{|{\bf r}|^{(2-n)/2}}{(2 \pi)^{n/2}} 
\int^{\infty}_0 \left( \sum^N_{j=1} c_j \lambda^{\alpha_j} \right)^{-1} 
\lambda^{n/2} \, J_{(n-2)/2} (\lambda |{\bf r}|) \, d \lambda,
\ee
where $n=1,2,3$ and $\alpha=(\alpha_1,...,\alpha_m)$.

These particular solutions allows us to describe
static fields in the elastic continuum with 
the weak spatial dispersion of $\alpha=(\alpha_1, . . . , \alpha_N)$-type.

\section{Fractional Weak Spatial Dispersion of $(\alpha,\beta)$-type}

\subsection{Fractional gradient elasticity equation for dispersion of $(\alpha,\beta)$-type}

If we have the dispersion law in the form
\be \label{a2c}
\hat{K}_{\alpha}(|{\bf k}|) \approx 
a_{\alpha} |{\bf k}|^{\alpha} + a_{\beta} |{\bf k}|^{\beta} + \hat{K}_{\alpha}(0)  ,
\ee
where $\alpha>1$, $\beta <3$, and $0<\beta < \alpha$, 
then we have the fractional gradient elasticity  equation 
\be \label{FPDE-4}
c_{\alpha} \, ((-\Delta)^{\alpha/2} u) ({\bf r}) + 
c_{\beta} \, ((-\Delta)^{\beta/2} u) ({\bf r}) = \frac{1}{\rho} \, f({\bf r}) ,
\ee
where
\be \label{cGj} 
c_{\alpha}= \frac{g \, a_{\alpha} \, d^{\alpha}}{M} , \quad 
c_{\beta}= \frac{g \, a_{\beta} \, d^{\beta}}{M} .
\ee
If $\alpha=4$ and $\beta=2$, we have the well-known equation of the gradient elasticity \cite{AA2011}:
\be \label{GradEl}
c_2 \, \Delta u ({\bf r}) - c_4 \Delta^2 u ({\bf r}) + \frac{1}{\rho} \, f({\bf r}) = 0,
\ee
where
\be \label{GradEl-2} 
c_2 = \frac{E}{\rho} = \frac{g \, a_2 \, d^2}{M} , \quad 
c_4 = \pm \, l^2 \, \frac{E}{\rho}= \frac{g \, a_4 \, d^4}{M} .
\ee
The second-gradient term is preceded by the sign that is defined by
$\operatorname{sgn} (g \, a_4)$, where $g \, a_2>0$.

Equation (\ref{FPDE-4}) is the fractional partial differential equation (\ref{FPDE-3}) with $n=3$, 
and such equation has the particular solution \cite{KST} of the firm
\be \label{phi-G4}
u({\bf r})=  \frac{1}{\rho} \, \int_{\mathbb{R}^3} 
G^3_{\alpha, \beta} ({\bf r} - {\bf r}^{\prime}) \, 
f({\bf r}^{\prime}) \, d^3 {\bf r}^{\prime},
\ee
where the Green type function is given by
\be \label{G-4}
G^3_{\alpha, \beta} ({\bf r}) =\frac{|{\bf r}|^{-1/2}}{(2 \pi)^{3/2}} 
\int^{\infty}_0 \left( c_{\alpha} \lambda^{\alpha}+ c_{\beta} |\lambda|^{\beta} \right)^{-1} 
\lambda^{3/2} \, J_{1/2} (\lambda |{\bf r}|) \, d \lambda .
\ee
Here $J_{1/2}$ is the Bessel function of the first kind.

\subsection{Point load problem for fractional gradient elasticity}

Let us consider point load problem for an infinite elastic continuum (see pages 25-26 in \cite{LL}), and
determine a deformation of an infinite gradient continuum,
when a force is applied to a small region in it.
We consider this Thomson's problem 
for non-local elastic continuum 
with fractional weak spatial dispersion of the form (\ref{a2c}). 
If we consider the deformation at distances $|{\bf r}|$, 
which are larger than the size of the region,
then we can assume that the force is applied at a point.
In this case, we have
\be \label{deltaf}
f({\bf r}) = f_0 \, \delta ({\bf r}) = f_0 \, \delta (x) \delta (y) \delta (z)  . 
\ee
Then the displacement field $u ({\bf r})$ of fractional gradient elasticity 
has a simple form of the particular solution (\ref{phi-G3}) 
that is proportional to the Green's function
\be \label{phi-Gb}
u ({\bf r}) = \frac{f_0}{\rho} \, G^n_{\alpha} ({\bf r}) ,
\ee
where $G^n_{\alpha}(z)$ is given by (\ref{G-3}). 
Therefore, the displacement field (\ref{phi-G4}) 
for the force that is applied at a point (\ref{deltaf}) has the form
\be \label{Pot-2}
u ({\bf r}) = \frac{1}{2 \pi^2} \frac{f_0}{ \rho \, |{\bf r}|} \, 
\int^{\infty}_0 \frac{ \lambda \, \sin (\lambda |{\bf r}|)}{ 
c_{\alpha} \lambda^{\alpha}+ c_{\beta} \lambda^{\beta}  } \, d \lambda .
\ee

From a mathematical point of view, there are two special cases:
(1) fractional weak spatial dispersion of $(\alpha,\beta)$-type with $\alpha=2$ and $0<\beta<2$;
(2) fractional weak spatial dispersion of $(\alpha,\beta)$-type with $\alpha \ne 2$, $\alpha > \beta$ and $0<\beta <3$.

From the point of view of the non-local elasticity theory 
is useful to distinguish two following particular cases:

\begin{itemize}

\item
Sub-gradient elasticity ($\alpha=2$ and $0<\beta<2$). 

\item
Super-gradient elasticity ($\alpha > 2$ and $\beta=2$). 

\end{itemize}

Note that for the first case the order of the fractional Laplacian 
less than the order of the first term related to the usual Hooke's law. 
In the second case the order of the fractional Laplacian greater 
of the order of the first term related to the Hooke's law.
The names of the sub- and super- gradient elasticity caused by the analogy 
with the names of anomalous diffusion \cite{MK1,Zaslavsky1,MK2}
such as subdiffusion and superdiffusion.


\subsection{Sub-gradient elasticity model}

The sub-gradient elasticity is characterized by the fractional weak spatial dispersion 
of $(\alpha,\beta)$-type with $\alpha=2$ and $0<\beta<2$.
Fractional model of non-local continuum with this spatial dispersion 
is described by equation (\ref{FPDE-4}) with $\alpha=2$ and $0<\beta<2$, given by 
\be \label{FPDE-4-2b1}
c_2 \Delta u ({\bf r}) - c_{\beta} ((-\Delta)^{\beta/2} u) ({\bf r}) +  \frac{1}{\rho} \, f({\bf r}) = 0 ,
\quad (0<\beta<2) .
\ee
The order of the fractional Laplacian $(-\Delta)^{\beta/2}$ less 
than the order of the first term related to the usual Hooke's law. 
As a simple example, consider the square of the Laplacian, i.e. $\beta=1$.

The particular solution of equation (\ref{FPDE-4-2b1})
for the force that is applied at a point (\ref{deltaf}) is the displacement field 
\be \label{Pot-2-2b}
u ({\bf r}) = \frac{1}{2 \pi^2} \frac{f_0}{\rho \, |{\bf r}|} \, 
\int^{\infty}_0 \frac{ \lambda \, \sin (\lambda |{\bf r}|)}{ 
c_2 \lambda^2+ c_{\beta} \lambda^{\beta}  } \, d \lambda.
\ee


Using equation (1) of Section 2.3 in the book \cite{BE}, 
we obtain the following asymptotic behavior for $u(|{\bf r}|)$ with $0<\beta<2$, when $|{\bf r}| \to \infty$
\be
u (|{\bf r}|) = \frac{f_0}{2\pi^2 \, \rho \, |{\bf r}|} \,
\int^{\infty}_0 \frac{\lambda \sin (\lambda |{\bf r}|)}{c_2 \lambda^2+ c_{\beta} \lambda^{\beta}}  \, d \lambda \approx
 \frac{ C_0(\beta) }{|{\bf r}|^{3-\beta}} + \sum^{\infty}_{k=1}  \frac{ C_k(\beta) }{|{\bf r}|^{(2-\beta)(k+1)+1}} ,
\ee
where 
\be
C_0(\beta)= \frac{f_0}{2\pi^2 \, \rho \, c_{\beta}} \, \Gamma(2-\beta) \, \sin \left( \frac{\pi}{2}\beta \right) ,
\ee
\be
C_k (\beta) = -\frac{f_0 c^k_2}{2 \pi^2 \,\rho \, c^{k+1}_{\beta}} \int^{\infty}_0  z^{(2-\beta)(k+1)-1} \, \sin(z) \, dz .
\ee

As a result, the displacement field for the force that is applied at a point
in the continuum with this type of fractional weak spatial dispersion is given by
\be
u ({\bf r}) \ \approx \ \, \frac{C_0 (\beta)}{|{\bf r}|^{3-\beta}} \quad (0< \beta<2)
\ee
on the long distance $|{\bf r}| \gg 1$.


\subsection{Super-gradient elasticity model}

The super-gradient elasticity is characterized by the fractional weak spatial dispersion 
of $(\alpha,\beta)$-type with $\alpha > 2$ and $\beta=2$.
For the non-local continuum with the weak spatial dispersion of the $(\alpha, \beta)$-type,
where $\alpha>\beta > 0$, $0<\beta<3$ and $\alpha \ne 2$
the displacement field for the fractional gradient model is described by equation
(\ref{FPDE-4}) includes two parameters $(\alpha, \beta)$. 
As an example of the non-local continuum with this type of spatial dispersion 
we highlight the case of super-gradient elasticity, where $\beta=2$ and $\alpha>2$.
In this case equation (\ref{FPDE-4}) has the form 
\be \label{FPDE-4-2b2}
c_2 \Delta u ({\bf r}) - c_{\alpha} ((-\Delta)^{\alpha/2} u) ({\bf r}) +  \frac{1}{\rho} \, f({\bf r}) = 0 ,
\quad (\alpha > 2) .
\ee
The order of the fractional Laplacian $(-\Delta)^{\alpha/2}$ 
greater of the order of the first term related to the Hooke's law.
If $\alpha =4$ equation (\ref{FPDE-4-2b2}) become the equation (\ref{GradEl}). 
Therefore the case $3<\alpha<5$ can be considered as close as possible ($\alpha \ \approx \ 4$)
to the usual gradient elasticity (\ref{GradEl}).

For the displacement field that is described by equation (\ref{FPDE-4}), 
where $\alpha>\beta > 0$, $0<\beta<2$, and $\alpha \ne 2$,
and the force $f({\bf r})$ is applied at a point (\ref{deltaf}),
we have the following asymptotic behavior 
\be
u (|{\bf r}|) \ \approx \  \frac{f_0 \, \Gamma(2-\beta) \sin(\pi \beta/2)}{2 \pi^2 \, \rho \, c_{\beta}} \,
\cdot \, \frac{1}{|{\bf r}|^{3-\beta}} \quad (|{\bf r}| \to \infty) .
\ee
We note that this asymptotic behavior $ |{\bf r}| \to \infty$ does not depend on the parameter $\alpha$.
The field on the long distances is determined only by term with $(-\Delta)^{\beta/2}$ ($\alpha>\beta$) 
that can be interpreted as a fractional non-local "deformation" of the Hooke's law. 

We note the existence of a maximum for the function
$u (|{\bf r}|) \cdot |{\bf r}|$ 
in the case $0<\beta < 2 < \alpha$. 

The asymptotic behavior of the displacement field $u (|{\bf r}|)$ for $|{\bf r}| \to 0$ is given by
\be \label{Cab-1}
u (|{\bf r}|) \ \approx \ 
\frac{f_0 \, \Gamma((3-\alpha)/2)}{ 2^{\alpha} \, \pi^2 \sqrt{\pi} \, \rho \, c_{\alpha} \, \Gamma(\alpha/2)} \,
\cdot \, \frac{1}{|{\bf r}|^{3-\alpha}} , \quad (1<\alpha<2),
\ee
\be \label{Cab-2}
u (|{\bf r}|) \ \approx \ 
\frac{f_0  \, \Gamma((3-\alpha)/2)}{2^{\alpha} \, \pi^2 \sqrt{\pi} \, \rho \, c_{\alpha} \, \Gamma(\alpha/2)} \,
\cdot \,  |{\bf r}|^{\alpha-3} , \quad (2<\alpha<3),
\ee
\be \label{Cab-3}
u (|{\bf r}|) \ \approx \ 
\frac{f_0}{2 \pi \, \alpha \, \rho \, c^{1-3/\alpha}_{\beta} \, c^{3/\alpha}_{\alpha} \, \sin (3 \pi / \alpha)}
 , \quad (\alpha>3),
\ee
where we use Euler's reflection formula for Gamma function.
The asymptotic relation (\ref{Cab-1}) is not directly related 
with the super-gradient case.  
Note that the above asymptotic behavior does not depend on the parameter $\beta$, 
and relations (\ref{Cab-1}-\ref{Cab-2}) does not depend on $c_{\beta}$.
The displacement field $u (|{\bf r}|)$ on the short distances 
is determined only by term with $(-\Delta)^{\alpha/2}$ ($\alpha>\beta$) 
that can be considered as a fractional non-local "deformation" of the gradient term.

\section{Conclusion}

A lattice model with spatial dispersion of
power-law type is suggested.
Gradient elasticity is considered as a phenomenological theory 
representing continuum limit of lattice dynamics,
where the length-scales are much larger than inter-atomic distances. 
In the continuum limit we derive continuum equations 
with spatial derivatives of non-integer order $\alpha$. 
The correspondent continuum equations
describe fractional generalization of 
gradient elasticity (the super-gradient elasticity model) for $\alpha>2$ and 
a special form of fractional integral 
elasticity (the sub-gradient elasticity model) for $0<\alpha <2$. 
The suggested lattice model with spatial dispersion 
can be considered as a microscopic basis for 
the fractional non-local elastic continuum.
We can note that a fractional nonlocal continuum model 
can be obtained from different microscopic or lattice models
\cite{CEPJ2013,MOM2014}. 
The main advantage of the suggested approach is that
we can use the Taylor series in the wave vector space
instead of Taylor expansion in a coordinate space.
It allows us to use these models 
as a microstructural basis of unified description 
of fractional (and integer) gradient models 
with positive and negative signs of the strain gradient terms.
The suggested approach can be generalized 
for three-dimensional case of gradient elasticity.
The proposed lattice model can also be easily generalized 
for the case of the high-order gradient elasticity 
and the correspondent fractional extension
by using the next terms of fractional Taylor series.
The suggested lattice models with long-range interactions
can be important to describe the non-local elasticity 
of materials at micro-scale and nano-scales \cite{Nano,Nano1,Nano2},
where the interatomic and intermolecular interactions 
are prevalent in determining the properties of these materials.


\section*{Acknowledgments}

The author expresses his gratitude to Professor Elias C. Aifantis for valuable discussions 
of fractional gradient elasticity and to Professor Juan J. Trujillo 
for valuable discussions of applications fractional models in elasticity theory.



\section*{Appendix A: Fractional Taylor Formula}

\subsection*{Riemann-Liouville and Caputo derivatives}

The left-sided Riemann-Liouville derivatives of order $\alpha >0$ are defined by
\be
(\, ^{RL}D^{\alpha}_{a+} f)(x) = \frac{1}{\Gamma(n-\alpha)} \left( \frac{d}{dx}\right)^n 
\int^x_a \frac{f(x^{\prime}) \, dx^{\prime}}{(x-x^{\prime})^{\alpha-n+1}} ,
\quad (n=[\alpha]+1) .
\ee
We can rewrite this relation in the form
\be
(\, ^{RL}D^{\alpha}_{a+} f)(x) =\left( \frac{d}{dx}\right)^n \, (I^{n-\alpha}_{a+} f)(x) ,
\ee
where $I^{\alpha}_{a+}$ is a  left-sided Riemann-Liouville integral of order $\alpha >0$ 
\be \label{RLI}
(I^{\alpha}_{a+} f)(x) = \frac{1}{\Gamma(\alpha)} 
\int^x_a \frac{f(x^{\prime}) \, dx^{\prime}}{(x-x^{\prime})^{1-\alpha}} ,
\quad (x>a) .
\ee

The Caputo fractional derivative of order $\alpha$ is defined by
\be
(\, ^CD^{\alpha}_{a+} f)(x) = \left( I^{n-\alpha}_{a+} \left( \frac{d}{dx}\right)^n f \right)(x) ,
\ee
where $I^{\alpha}_{a+}$ is a  left-sided Riemann-Liouville integral (\ref{RLI}) of order $\alpha >0$.
In equation (\ref{Tay-Cap}) we use $0<\alpha<1$ and $n=1$. 
The main distinguishing feature of the Caputo fractional derivative is that, 
like the integer order derivative, 
the Caputo fractional derivative of a constant is zero. 

Note also that the third term in (\ref{Tay-Cap}) involves the fractional derivative 
of the a fractional derivative, which is not the same as the $2 \alpha$ fractional
derivative. In general,
\be 
(\, ^CD^{\alpha}_{a+} \, ^CD^{\alpha}_{a+} f)(x) \ne (\, ^CD^{2\alpha}_{a+} f)(x) . 
\ee
Then the coefficients of the fractional Taylor
series can be found in the usual way, by repeated differentiation.
This is to ensure that the fractional derivative of order $\alpha$ 
of the function $(x-a)^{\alpha}$ is a constant. The repeated 
fractional derivative of order $\alpha$ gives zero. 
Then the coefficients of the fractional Taylor
series can be found in the usual way, by repeated differentiation.

\subsection*{Fractional Taylor's series in the Riemann-Liouville form}

Let $f(x)$ be a real-value function such that 
the derivative $(\, ^{RL}D^{\alpha+m}_{a+} f)(x)$ is integrable. 
Then the following analog of Taylor formula holds 
(see Chapter 1. Section 2.6 \cite{SKM}):
\be
f(x)= \sum^{m-1}_{j=0} \frac{(\, ^{RL}D^{\alpha+j}_{a+}f)(a+)}{\Gamma(\alpha+j+1)} \, 
(x-a)^{\alpha+j} +R_m (x) , \quad (\alpha>0) ,
\ee
where $D^{\alpha+j}_{a+}$ are left-sided Riemann-Liouville derivatives, and
\be
R_m (x) = (I^{\alpha+m}_{a+} \, ^{RL}D^{\alpha+m}_{a+} f)(x) .
\ee

\subsection*{Riemann formal version of the generalized Taylor's series}

The Riemann formal version of the generalized Taylor's series \cite{Riem,Hardy}:
\be
f(x)= \sum^{+\infty}_{m=-\infty} \frac{(\, ^{RL}D^{\alpha+m}_a f)(x_0)}{\Gamma(\alpha+m+1)} (x-x_0)^{\alpha+m} ,
\ee
where $\, ^{RL}D^{\alpha}_a$ for $\alpha>0$ is the Riemann-Liouville fractional derivative, and
$\, ^{RL}D^{\alpha}_a=I^{-\alpha}_a$ for $\alpha<0$ is the Riemann-Liouville fractional integral of order $|\alpha|$.

\subsection*{Fractional Taylor's series in the Trujillo-Rivero-Bonilla form}

The Trujillo-Rivero-Bonilla form of generalized Taylor's formula \cite{TRB} :
\be
f(x) = \sum^{m}_{j=0} \frac{c_j}{\Gamma((j+1)\alpha)} \, (x- a)^{(j+1)\alpha-1}  +R_m(x,a),
\ee
where $\alpha \in [0;1]$, and
\be
c_j = \Gamma(\alpha)  \, [(x-a)^{1-\alpha} \, (\, ^{RL}D^\alpha_a)^j f(x)](a+) ,
\ee
\be
R_m(x,a) = \frac{ ((\, ^{RL}D^{\alpha}_a)^{m+1} f)(\xi) }{ \Gamma((m+1) \alpha+1)} \,
 (x-a)^{(m+1)\alpha} , \quad \xi \in [a;x] .
\ee

\subsection*{Fractional Taylor's series in the Dzherbashyan-Nersesian form}

Let $\alpha_k$, $(k=0,1,...,m)$ be increasing sequence of real numbers such that
\be
0 < \alpha_k-\alpha_{k-1} \le 1, \quad \alpha_0=0, \quad k=1,2,...,m.
\ee

We introduce the notation \cite{Arm1,Arm2} (see also Section 2.8 in \cite{SKM}):
\be
D^{(\alpha_k)}  = I^{1-(\alpha_k-\alpha_{k-1} )}_{0+} D^{1+ \alpha_{k-1} }_{0+} .
\ee
In general, $D^{(\alpha_k)} \ne ^{RL}D^{\alpha_k}_{0+} $. 
Fractional derivative $D^{(\alpha_k)}$ differs from the Riemann-Liouville derivative 
$^{RL}D^{\alpha_k}_{0+} $ by finite sum of power functions since (see Eq. 2.68 in \cite{KST})
\be
I^{\alpha}_{0+} I^{\beta}_{0+} \ne I^{\alpha+\beta}_{0+} .
\ee
 
The generalized Taylor's formula \cite{Arm1,Arm2}
\be
f(x) = \sum^{m-1}_{k=0} a_k \, x^{\alpha_k} +R_{m}(x), \quad (x>0).
\ee
where
\be
a_k = \frac{(D^{(\alpha_k)}  f)(0)}{\Gamma(\alpha_k+1)} , \quad
R_{m}(x) = \frac{1}{\Gamma(\alpha_m+1)} \int^x_0 (x-z)^{\alpha_m-1} \, (D^{(\alpha_k)}  f)(z) \, dz .
\ee

\subsection*{Fractional Taylor's series in the Odibat-Shawagfeh form}

The fractional Taylor series is a generalization of the Taylor series 
for fractional derivatives, where $\alpha$ is the fractional
order of differentiation, $0<\alpha<1$. 
The fractional Taylor series with Caputo derivatives \cite{OdbSh}
has the form
\be \label{Tay-Cap}
f(x)=f(a)+ \frac{(\, ^CD^{\alpha}_{a+}f)(a)}{\Gamma (\alpha+1)} (x-a)^{\alpha}
+  \frac{(\, ^CD^{\alpha}_{a+} \, ^CD^{\alpha}_{a+} f)(a)}{\Gamma (2\alpha+1)} (x-a)^{2 \alpha}+ . . . ,
\ee
where $\, ^CD^{\alpha}_{a+}$ is the Caputo fractional derivative of order $\alpha$.


\section*{Appendix B: Riesz fractional derivatives and integrals}

Fractional integration and fractional differentiation 
in the $n$-dimensional Euclidean space $\mathbb{R}^n$ 
can be defined as fractional powers of the Laplace operator. 
For $\alpha > 0$ and "sufficiently good" functions $f(x)$, 
$x \in \mathbb{R}^n$, the fractional Laplacian in the Riesz's form
(the Riesz fractional derivative)
is defined in terms of the Fourier transform ${\cal F}$ by
\be \label{RFD-1}
((-\Delta)^{\alpha/2} f)(x)= {\cal F}^{-1} \Bigl( |k|^{\alpha} ({\cal F} f)(k) \Bigr) .
\ee
The Riesz fractional integration is defined by
\be
{\bf I}^{\alpha}_x f(x) = 
{\cal F}^{-1} \Bigl( |k|^{-\alpha} ({\cal F} f)(k) \Bigr) .
\ee

The Riesz fractional integration can be realized in the form of 
the Riesz potential defined as the Fourier's convolution of the form
\be
{\bf I}^{\alpha}_x f(x)= 
\int_{\mathbb{R}^n} K_{\alpha}(x-z) f(z) dz, \quad (\alpha >0) ,
\ee
where the function $K_{\alpha}(x)$ is the Riesz kernel.
If $\alpha>0$, and $\alpha \not=n,n+2,n+4,...$, 
the function $K_{\alpha}(x)$ is defined by 
\be 
K_{\alpha}(x) = \gamma^{-1}_n(\alpha) |x|^{\alpha-n} . 
\ee
If $\alpha \not=n,n+2,n+4,...$, then
\be 
K_{\alpha}(x) = - \gamma^{-1}_n(\alpha) |x|^{\alpha-n} \ln |x| . \ee
The constant $\gamma_n(\alpha)$ has the form
\be \gamma_n(\alpha)=
\begin{cases}
2^{\alpha} \pi^{n/2}\Gamma(\alpha/2)/ \Gamma(\frac{n-\alpha}{2}) &
\alpha \not=n+2k, \quad n \in \mathbb{N},
\cr
(-1)^{(n-\alpha)/2}2^{\alpha-1} \pi^{n/2} \;
\Gamma(\alpha/2) \; \Gamma( 1+[\alpha-n]/2) 
 & \alpha =n+2k.
\end{cases}
\ee

Obviously, the Fourier transform of the Riesz fractional integration is given by
\be 
{\cal F} \Bigl( {\bf I}^{\alpha}_x f(x)\Bigr) = 
|k|^{-\alpha} ({\cal F} f)(k) . 
\ee
This formula is true for functions $f(x)$ belonging to Lizorkin's space. 
The Lizorkin spaces of test functions on $\mathbb{R}^n$
is a linear space of all complex-valued infinitely differentiable
functions $f(x)$ whose derivatives vanish at the origin:
\be
\Psi=\{ f(x): f(x) \in S(\mathbb{R}^n), \quad 
(D^{\bf n}_xf)(0)=0, \quad |{\bf n}| \in \mathbb{N} \} ,
\ee
where $S(\mathbb{R}^n)$ is the Schwartz test-function space.
The Lizorkin space is invariant with respect 
to the Riesz fractional integration. 
Moreover, if $f(x)$ belongs to the Lizorkin space, then
\be
 {\bf I}^{\alpha}_x \, {\bf I}^{\beta}_x f(x) 
= {\bf I}^{\alpha+\beta}_x f(x) , 
\ee
where $\alpha >0$, and $\beta >0$.


For $\alpha >0$, the 
the fractional Laplacian in the Riesz's form
can be defined in the form of the hypersingular integral by
\be 
((-\Delta)^{\alpha/2}f)(x) =
\frac{1}{d_n(m,\alpha)} \int_{\mathbb{R}^n} 
\frac{1}{|z|^{\alpha+n}} (\Delta^m_z f)(z) \, dz , 
\ee
where $m> \alpha$, and $(\Delta^m_z f)(z)$ is a finite difference of
order $m$ of a function $f(x)$ with a vector step $z \in \mathbb{R}^n$
and centered at the point $x \in \mathbb{R}^n$:
\be 
(\Delta^m_z f)(z) =
\sum^m_{k=0} (-1)^k \frac{m!}{k!(m-k)!}  \, f(x-kz) . 
\ee
The constant $d_n(m,\alpha)$ is defined by
\be d_n(m,\alpha)=\frac{\pi^{1+n/2} A_m(\alpha)}{2^{\alpha} 
\Gamma(1+\alpha/2) \Gamma(n/2+\alpha/2) \sin (\pi \alpha/2)} ,  
\ee
where
\be 
A_m(\alpha)=\sum^m_{j=0} (-1)^{j-1} 
\frac{m!}{j!(m-j)!} \, j^{\alpha} . 
\ee
Note that the hypersingular integral $((-\Delta)^{\alpha/2}f)(x)$ 
does not depend on the choice of $m>\alpha$.

If $f(x)$ belongs to the space of "sufficiently good" functions, then
the Fourier transform ${\cal F}$ of the fractional Laplacian in the Riesz's form
is given by 
\be 
({\cal F} (-\Delta)^{\alpha/2} f)(k) = |k|^{\alpha} ({\cal F}f)(k) . 
\ee
This equation is valid for the Lizorkin space \cite{SKM}
and the space $C^{\infty}(\mathbb{R}^n)$ of infinitely differentiable 
functions on $\mathbb{R}^n$ with compact support.

The the fractional Laplacian in the Riesz's form yields an operator inverse 
to the Riesz fractional integration for a special space of functions.
The formula
\be \label{bfDa}
(-\Delta)^{\alpha/2} \, {\bf I}^{\alpha}_x f(x) = f(x) , \quad (\alpha >0) \ee
holds for "sufficiently good" functions $f(x)$. 
In particular, equation (\ref{bfDa}) for $f(x)$ belongs to the Lizorkin space.
Moreover, this property is also valid for the Riesz fractional integration in
the frame of $L_p$-spaces: $f(x) \in L_p(\mathbb{R})$ for$1 \leqslant p < n/a$ 
(see Theorem 26.3 in \cite{SKM}).


\end{document}